\documentclass[aps,prc,preprint,showpacs,amsmath,amssymb,nofootinbib]{revtex4}
\usepackage {lscape}
\usepackage{graphicx}
\usepackage{amsmath}

\begin{document}

\title{High-spin structure of $^{87}$Sr and $^{87}$Zr nuclei: shell model interpretation}
\author{P.C. Srivastava}
\thanks{pcsrifph@iitr.ac.in}
\address{Department of Physics, Indian Institute of Technology, Roorkee - 247 667, India}

\author{Vikas Kumar}
\address{Department of Physics, Indian Institute of Technology, Roorkee - 247 667, India}

\author{M. J. Ermamatov}
\address{Instituto de F\'isica, Universidade Federal Fluminense, 24210-340, Niter\'oi, Rio de Janeiro, Brazil}
\address{Institute of Nuclear Physics, Ulughbek, Tashkent  100214, Uzbekistan}

\date{\hfill \today}
\begin{abstract}

 In the present work we report a comprehensive analysis of shell model results for high-spin states 
of  $^{87}$Sr and $^{87}$Zr for recently available experimental data within the full
$f_{5/2}pg_{9/2}$ model space using JUN45 and
jj44b effective interactions developed
for this  model space.  In this work we have compared the energy levels, electromagnetic transition probabilities, quadrupole and magnetic moments
with available experimental data. We have confirmed structure of high-spin states of these two nuclei which were tentatively assigned in the recent experimental work.
In the case of $^{87}$Sr, for positive parity states up to $\sim$ 7.5 MeV, both interactions  
predict very good agreement with experimental data,  while negative parity states are slightly suppressed in jj44b calculation.
For the $^{87}$Zr nucleus, the jj44b interaction predicts higher energies for the negative parity states beyond $J \geq 27/2^{-}$.
The configuration, which have one hole in $\nu g_{9/2}$ orbital, is responsible for generating the states in $^{87}$Sr. In the case of $^{87}$Zr,
low-lying positive parity states come with the configuration having three holes in the $\nu g_{9/2}$, while the odd-parity states have configuration 
$\nu (f_{5/2}^{-1}g_{9/2}^{-2})$.
\end{abstract}


\maketitle

\section{Introduction}

Experimental information on low-lying single-particle excited states are available for many nuclei.  Because of the advancement in experimental techniques now it is possible to populate high spin states of nuclei beyond Ni.
These experimental results are stringent test for the predicting power of shell model for high spin excited states.
It is possible to test two-body matrix elements for predicting the high-spin states which are generated by aligning
the angular momenta due to broken nucleon pairs. The nuclei in Sr-Zr region show many interesting features, such
as spherical shell, isomeric states, candidates of double and neutrinoless double beta decay \cite{Kum,th1,jun45,ecl,th2,th3,th4,th5,th6,th7,th8,jain}.

High spin states of $^{88}$Zr  have been recently  populated up to $\sim$ 20$\hbar$ and an excitation energy of ~10 MeV was measured in $^{80}$Se($^{13}$C,5n)$^{88}$Zr fusion evaporation reaction~\cite{zr88}.  
Similarly, the high spin states of $^{89}$Zr were populated using the fusion evaporation reaction $^{80}$Se($^{13}$C,$4n$)$^{89}$Zr.  The observed high-spin states up to 10 MeV excitation energy and spin $\sim$ 37/2$\hbar$ are reported in Ref.~\cite{zr89}. The dominance of single-particle excitations is shown for both positive and negative parity states.
The high-spin band structure of $^{85}$Sr was populated in the reaction $^{76}$Ge($^{13}$C,$4n$)$^{85}$Sr \cite{sr85}. The spin and parity of different levels up to the spin of $\sim$ 35/2$\hbar$ and an excitation energy $\sim$ 7.5 MeV were established.
Here shell model explains various features, such as the odd-even staggering, very well. In the case of $^{86}$Sr 
the high-spin states were populated using $^{76}$Ge($^{13}$C,$3n$)$^{86}$Sr reaction. 
The level scheme up to 10.9 MeV excitation energy and
maximum spin of $\sim$ 19$\hbar$ have been reported in Ref.~\cite{sr86}. 

Experimentally, the high-spin structure of $^{87}$Sr previously was studied in Refs.~\cite{sr871,sr872}. Recently, using the fusion-evaporation reaction 
$^{82}$Se($^{9}$Be,$4n$)$^{87}$Sr, the states were populated up to an excitation energy of 7.4 MeV at 
spin 31/2$\hbar$ reported in Ref.~\cite{sr873}. The structure of high spin states using in-beam $\gamma$-ray spectroscopic method $^{87}$Zr was studied through the $^{59}$Co($^{32}$S,$3pn$)$^{87}$Zr
reaction \cite{zr871}, the  level scheme was established up to spin (37/2$^+$) and (43/2$^-$).

Motivated by the success of our shell model results in this region for recently measured high spin states of $^{88}$Zr\cite{zr88}, $^{89}$Zr~\cite{zr89}, 
$^{85}$Sr~\cite{sr85} and $^{86}$Sr~\cite{sr86},
in the present work we will be focusing on the shell model study of recently populated high-spin states of $^{87}$Sr and $^{87}$Zr.

\section{Shell model calculation}

In the present shell model calculations $^{56}$Ni is taken as the inert core
with the spherical orbits $1p_{3/2}$, $0f_{5/2}$, $1p_{1/2}$ and $0g_{9/2}$.
We have performed calculation with the jj44b 
and JUN45 effective interactions. The jj44b interaction was fitted with 600
binding energies and excitation energies from nuclei with $Z = 28-30$ and $N =
48-50$ available in this region. Here, 30 linear combinations of $JT$ coupled two-body matrix elements
(TBME) are varied, giving the rms deviation of about 250 keV from  the experiment.
The single particle energies (spe) are taken to be  -9.656,
-9.287, -8.269 and -5.894 MeV for the $p_{3/2}$, $f_{5/2}$, $p_{1/2}$ and
$g_{9/2}$ orbitals, respectively \cite{jj44b}. For the
JUN45, the single-particle energies for the 1$p_{3/2}$,
0$f_{5/2}$, 1$p_{1/2}$ and 0$g_{9/2}$ orbitals are -9.828, -8.709, -7.839, and -6.262 MeV,
respectively. The JUN45 \cite{jun45} interaction is based on Bonn-C potential, the  single-particle energies and
two-body matrix elements were modified empirically with A = 63$\sim$69 mass region.
We have performed calculations 
using the shell model code Antoine \cite{antoine}. The maximum 
matrix dimension in M-scheme  $>$ $30$ millions for $^{87}$Zr.  

\subsection{Shell model results for $^{87}$Sr}

Experimental data are available from the earlier works~\cite{sr871,sr872} and recent work~\cite{sr873}, where the states are populated up to the excitation energy of 7.4 MeV with 31/2$\hbar$ spin. Useful structural information can be extracted through the study of this nucleus since both number of protons and neutrons are near closed shells and in this region spherical and collective behaviors of nuclei are important.
 In Fig.~1 the comparison of our shell model calculation with the experimental data are shown, where we have used two different JUN45 and jj44b interactions.  

The ground state $9/2_1^+$, which comes from the $\nu (g_{9/2}^{-1})$ configuration, is predicted correctly by both calculations.     
In the JUN45 calculation the values of $5/2^+_1$ and $7/2^+_1$ energy levels are only 65 keV lower and 3 keV higher than in the experiment, respectively.
In the jj44b calculation they are 86 keV lower and 351 keV higher than in the experiment, respectively.
The $1/2^+_1$, $3/2^+_1$, $5/2^+_2$ and $9/2^+_2$ levels are closer to the experiment in the jj44b calculation.
The $3/2^+_2$ level is far from experiment almost at the same amount in the both calculations.  The $13/2^+_1$ level comes from the
$\pi (f_{5/2}^{-1}p_{1/2}^{1})\otimes\nu (g_{9/2}^{-1})$ configuration. It is predicted 163 keV and 178 keV lower than in the experiment in the JUN45 and jj44b calculations, respectively.

As is seen from Fig.~1 all the $17/2_1^+$ - $31/2_1^+$ levels are predicted lower than in the experiment in the JUN45 calculation. In the jj44b
calculation these levels are little bit lower than even in the JUN45 calculation up to $27/2_1^+$, but $29/2_1^+$ and $31/2_1^+$ levels are well predicted by this calculation.       
According to the both shell model calculations these states come from $\pi ((p_{3/2}f_{5/2}p_{1/2})^{-2}(g_{9/2}^{2}))\otimes\nu (g_{9/2}^{-1})$ 
configuration. Probabilities vary within ~35\% - ~60\% and  ~24\% - ~49\% in the calculations with the JUN45 and jj44b interactions, respectively.
The JUN45 calculation predicts  $17/2_2^+$ and $25/2_2^+$ levels in the experiment only  with 71 keV  and with 279 keV differences, respectively.
In the jj44b calculation they are comparatively low.   
  
The negative parity $1/2^-$, $3/2^-$, $5/2^-$, $9/2^-_1$ and $11/2^-_1$ levels, which are due to the neutrons in $pf$ shell, are better predicted by the JUN45 calculation. The $7/2^-$ level which appears in both calculations is not measured yet in the experiment, although experimentally the doublet of levels ($5/2^-$,$7/2^-$)  is supposed to be at 2656 keV.
The negative parity $13/2_1^-$ - $23/2_1^-$ states are due to  
$\pi ((p_{3/2}f_{5/2}p_{1/2})^{-1}(g_{9/2}^{1}))\otimes\nu (g_{9/2}^{-1})$ configuration, with ~35\% - ~64\% and ~33\% - ~53\% probabilities, in the calculations with JUN45 and jj44b interactions, respectively. As is seen from Fig.~1 though the distance between $13/2^-$ and $15/2^-$ is larger and those of between the pair of levels $17/2^-$ and $19/2^-$, and $21/2^-$ and $23/2^-$ are more compressed, in general the negative parity states are described reasonably well by JUN45 calculation.  In jj44b calculation all negative parity levels, except $19/2_2^-$, are lower than in the JUN45 calculation. 
    
\begin{figure}
\centering
\includegraphics[scale=1.0]{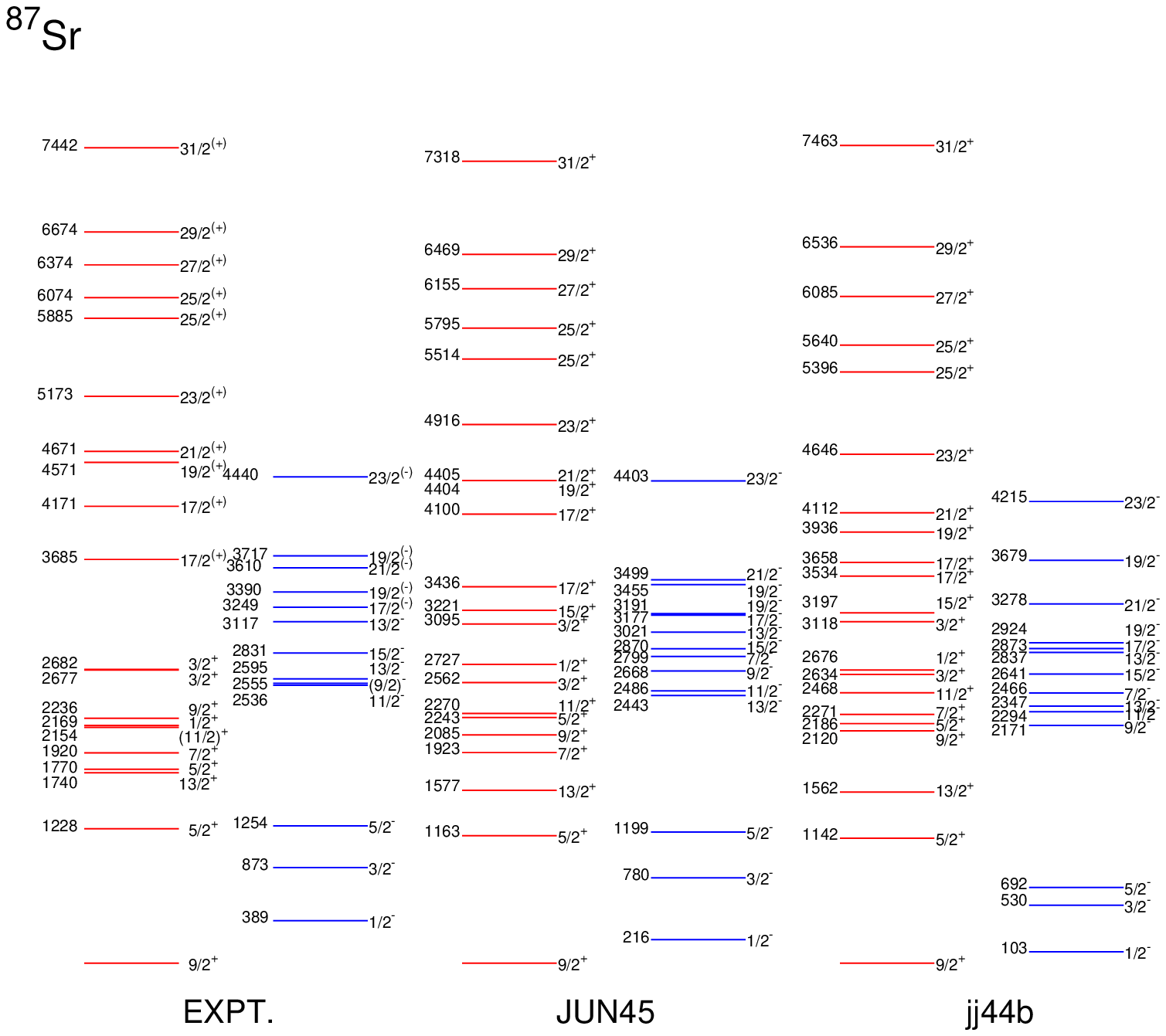}
\caption{Comparison of the theoretical and experimental energy levels of the $^{87}$Sr.} 
\label{sr} 
\end{figure}
\subsection{Shell model results for $^{87}$Zr}

The structure of high spin states of $^{87}$Zr was studied using in-beam $\gamma$-ray spectroscopic method through the $^{59}$Co($^{32}$S,3pn)$^{87}$Zr \cite{zr871}.  Positive  parity level scheme was established up to spin (37/2$^+$) and the negative parity level scheme up to (43/2$^-$). 

As in the case of $^{87}$Sr ground state spin of the $^{87}$Zr is also $9/2^+$ since still neutrons in the $g_{9/2}$ orbital play major role for the ground state. Now we see that less energy is needed in order to excite nuclei to $13/2^+_1$ state. This is reasonable
since now neutrons are little bit further from the filling  $g_{9/2}$ orbital as compared to the $^{87}$Sr nucleus.  In the calculation with
JUN45 interaction the $9/2_1^+$, $13/2_1^+$, $11/2_1^+$, $21/2_1^+$, $25/2_1^+$, $29/2_1^+$, $31/2_1^+$, $33/2_1^+$, $35/2_1^+$ and $37/2_1^+$ states
have the configuration $\pi (g_{9/2})^{2}\otimes\nu (g_{9/2})^{-3}$ with probabilities ~12\% - ~31\%.
The $7/2_1^+$, $17/2_1^+$ and $29/2_2^+$ states have the configuration $\pi [(p_{3/2}f_{5/2}p_{1/2})^{-2}(g_{9/2})^{2}]\otimes\nu (g_{9/2})^{-3}$.
Here the lower proton orbitals contribute to the configuration of these states.
In the calculation with jj44b interaction, the $9/2_1^+$, $7/2_1^+$, $13/2_1^+$, $11/2_1^+$, $25/2_1^+$ and $29/2_2^+$ states have 
$\pi [(p_{3/2}f_{5/2})^{-2}(g_{9/2})^{4}]\otimes\nu (g_{9/2})^{-3}$ configuration. 

From Fig.~2 we can see that between the pair of levels $9/2_1^+$ and $13/2_1^+$ there are $7/2_1^+$ and $7/2_2^+$ levels and between
the $13/2_1^+$ and $17/2_1^+$ levels there are $11/2_1^+$ and $11/2_2^+$ levels in the experiment.  In  both calculations the $5/2_1^+$ and
$7/2_1^+$ levels appear between the $9/2_1^+$ and $13/2_1^+$ levels, and the $7/2_2^+$ level appears after the $11/2_1^+$ level. In Fig.~2 we have not
shown the levels for which spins are not assigned in the experiment. But in the experiment there exist spin not assigned levels with energies 523.7 and 589.7 keV~\cite{nndc}
which are close to the calculated $5/2^+_1$ level.  
The $7/2_1^+$ is close to the experiment in jj44b, but the $7/2_2^+$ level is much higher in both calculations. In the JUN45 calculation
$11/2_1^+$ and $11/2_2^+$ levels are between $13/2_1^+$ and $17/2_1^+$ levels as in the experiment, but in jj44b calculation $11/2_2^+$ level
is located after $17/2_1^+$ level. The $1/2_1^+$, $3/2_1^+$, $9/2_2^+$, and $19/2_1^+$ levels can be spin not assigned levels observed
in the experiment~\cite{nndc}.  The levels $21/2_1^+$ and $25/2_1^+$ are only 22 and 79 keV higher than in the experiment in the jj44b
calculation, respectively. Between these levels there are $19/2_1^+$ and $23/2_1^+$ levels which also appear in the JUN45 calculation and are not
measured in the experiment.  In the JUN45 calculation between the levels $25/2_1^+$ and $29/2_1^+$, which are lower than in the experiment in both
calculations, there are $25/2^+_2$ and $27/2^+_1$  levels. They are not measured in the experiment. In the jj44b calculation
only $27/2_1^+$ level appears between the levels $25/2_1^+$ and $29/2_1^+$. The sequence of levels $29/2_1^+$ and $29/2_2^+$ are the same with the
experimental one in both JUN45 and jj44b calculations.  Though the distance between the levels are quite similar to the
experiment, in the JUN45 calculation they are little bit lower than in the experiment. In the jj44b calculation the first of these levels is only 79 keV higher and the second
one is 208 keV lower than in the experiment and the distance between the levels is little bit compressed as compared to the experiment and the JUN45 calculation.

\begin{figure*}
\centering
\includegraphics[scale=1.0]{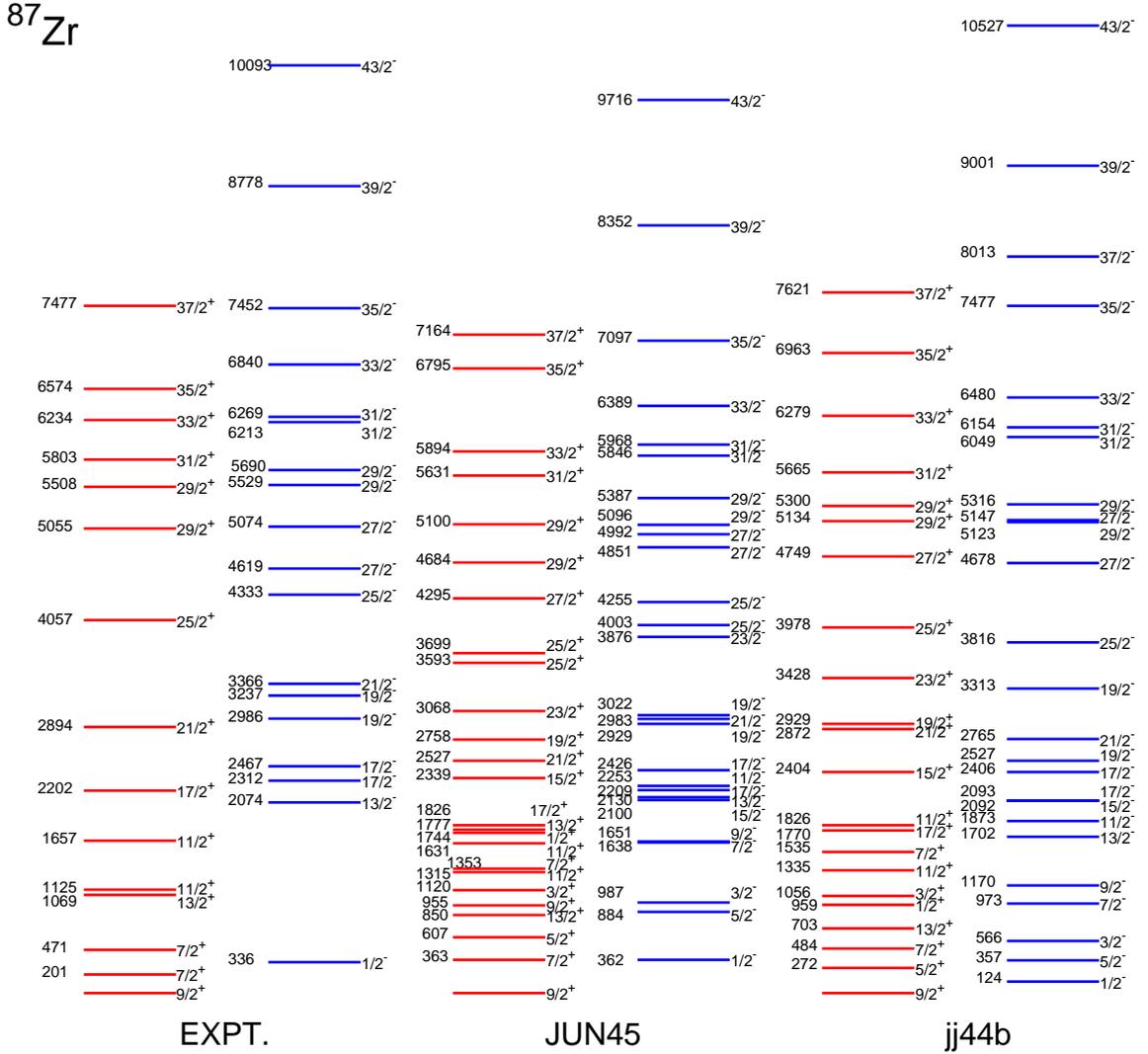}
\caption{ Comparison of the theoretical and experimental energy levels of the $^{87}$Zr.}  
\label{zr} 
\end{figure*}

For  $I\ge29/2_1^+$ states the $\pi [(p_{3/2}f_{5/2})^{-1}(g_{9/2})^{3}]\otimes\nu [(f_{5/2}^{-1}g_{9/2}^{-2}]$ configuration dominates. Agreement of the calculated $31/2^+$ level with the experimental one is approximately the same in both calculations. However the $33/2^+$ level is in better agreement with the experimental one in the jj44b calculation and is lower in the JUN45 calculation.  The $35/2^+$ level is located higher than in the experiment in both calculations. In the JUN45 calculation its value is closer to the experimental one than in the jj44b calculation. The $37/2^+$ level is predicted lower than in the experiment in the JUN45 calculation and is predicted higher in the jj44b calculation.  

The arrangement of lowest negative parity levels of $^{87}$Zr are very similar to those of $^{87}$Sr. As in the case of $^{87}$Sr we have not shown the levels for which spins are not assigned in the experiment. Therefore $3/2^-$,  $5/2^-$, $7/2^-$ and $9/2^-$ levels which appear in the calculations may be one of this spin not assigned levels. Obviously these  levels are due to the neutrons in $pf$ shell. In jj44b the $1/2^-$ level is predicted much lower than in the experiment. In the JUN45 calculation it is predicted only 26 keV higher than in the experiment.  

The $13/2_1^-$, $21/2_1^-$, $25/2_1^-$, $29/2_1^-$, $29/2_2^-$, $31/2_1^-$ and $33/2_1^-$  levels come from $\pi (g_{9/2}^{2})\otimes\nu (f_{5/2}^{-1}g_{9/2}^{-2})$ configuration with ~6\% - ~18\% probabilities, respectively. All these levels are lower than in the experiment in both calcilations. The $13/2_1^-$, $21/2_1^-$, $25/2_1^-$ levels are better described by the JUN45 interaction. The quality of agreement of $29/2_1^-$ and $29/2_2^-$ levels is more or less the same in both calculations.  The $31/2_1^-$ and $33/2_1^-$ levels are closer to the experiment in the calculation with jj44b interaction. 

The $17/2_2^-$  state comes from $\pi (g_{9/2}^{2})\otimes\nu (p_{1/2}^{-1})$ 
configuration which is close to the experiment in the JUN45 calculation and the jj44 calculation differs from the JUN45 only to 20 keV. 

The states $17/2_1^-$, $19/2_1^-$, $19/2_2^-$, $27/2_1^-$, and $31/2_2^-$  levels come from $\pi (p_{1/2}^{1}g_{9/2}^{1})\otimes\nu (g_{9/2}^{-3})$ configuration with ~16\% - ~38\% probabilities, respectively. The $17/2_1^-$, $19/2_1^-$ and $27/2_1^-$ levels are in better agreement in the JUN45 calculation, while $19/2_2^-$, and $31/2_2^-$ are better in the jj44b calculation.

The $27/2_2^-$, $35/2_1^-$, and $39/2_1^-$ levels come from $\pi (f_{5/2}^{-1}g_{9/2}^{3})\otimes\nu (g_{9/2}^{-3})$ configuration  with ~24\% - ~50\% probabilities, respectively. These states are described well by the calculation with jj44b interaction.  

The experimental $43/2^-$ level is measured at 10093 keV. The JUN45 and jj44b calculations predict this level at 9716 keV  and at 10527 keV, respectively.

\begin{figure}
\centering
\begin{center}
$^{87}$Sr 
\end{center}
\includegraphics[scale=0.65]{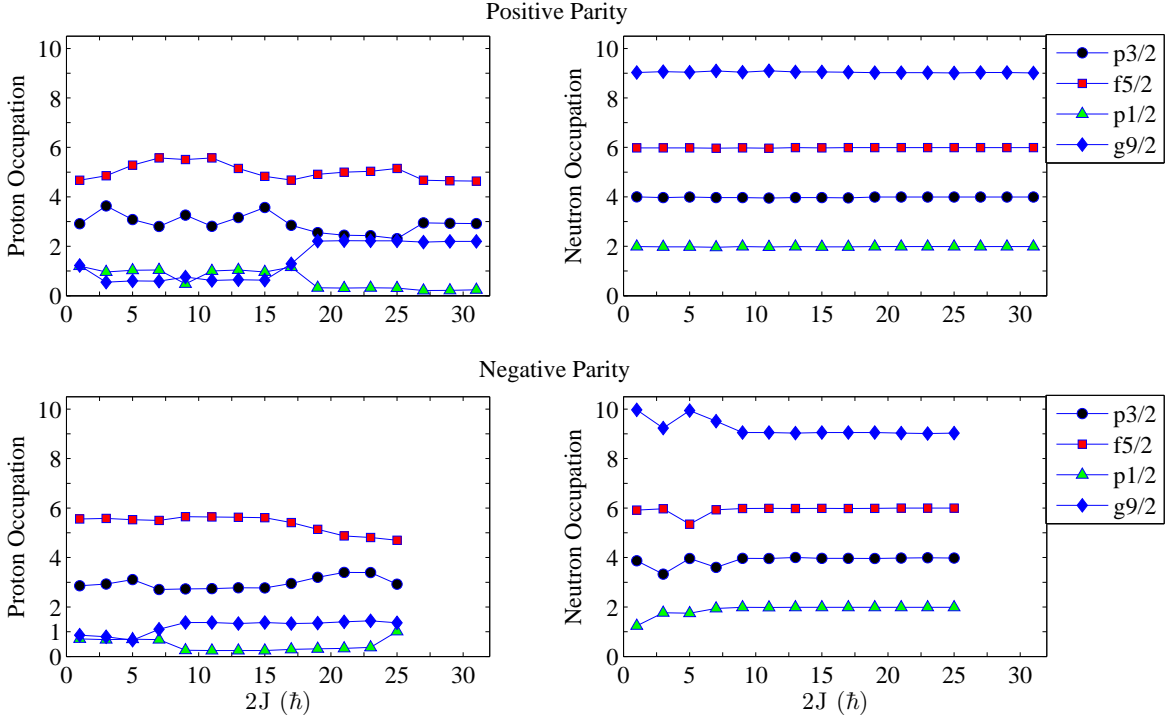}
\caption{Occupancy of different protons and neutrons orbitals with JUN45 interaction for $^{87}$Sr.}  
\label{srzr_occ} 
\end{figure}
\subsection{Occupancy of the orbitals}
In order to look closely to the structure of the states  we show the occupancy of different protons
and neutrons orbitals for $^{87}$Sr and $^{87}$Zr nuclei with JUN45 interaction in Figs.~3 and 4, respectively.
As is seen from Fig.~3, for the positive parity states in $^{87}$Sr the  occupancy of the proton orbitals are sensitive
to the nuclear spin, including states up to high spins. Here the $\pi g_{9/2}$ occupancy is increasing at the expense
of $\pi f_{5/2}$ and  $\pi p_{1/2}$ occupancy. For the negative parity states the dependence of the  occupancy of the
proton orbitals from the spins still remains same, however now  the neutron occupancy at lower spins shows irregular pattern.
For negative parity states the occupancy of $\pi g_{9/2}$ is increasing at the expense of the $\pi p_{1/2}$ orbital occupancy only.

From Fig.~4 one can see that  the proton occupancy becomes more stable up to high spins in the positive parity states
of the $^{87}$Zr nucleus as compared to the $^{87}$Sr nucleus. The visible changes in the occupancy with respect to spins
can be seen in $g_{9/2}$ orbital, the increase in the occupancy is at the expense of the $\pi p_{1/2}$ orbital occupancy only, as in the case of $^{87}$Sr.   
\begin{figure}
\begin{center}
$^{87}$Zr 
\end{center}
\includegraphics[scale=0.65]{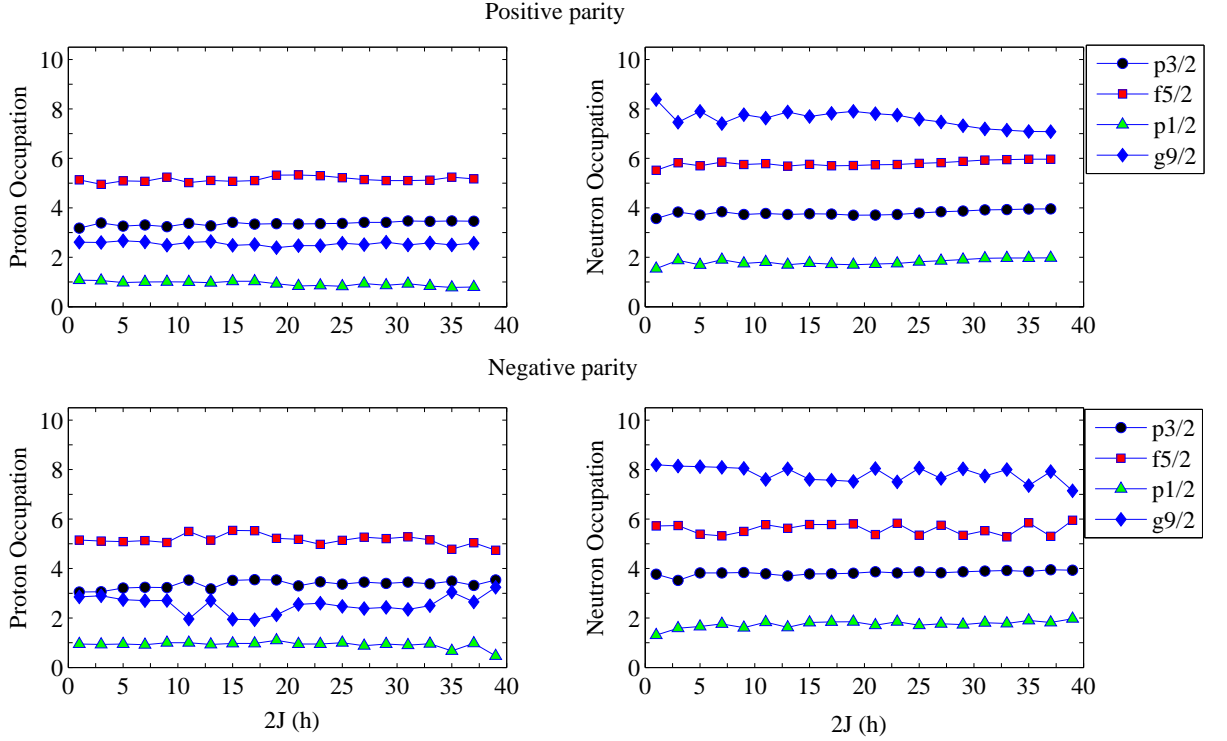}
\caption{Occupancy of different protons and neutrons orbitals with JUN45 interaction for $^{87}$Zr.}  
\label{srzr_occ} 
\end{figure}

\begin{table}[hbtp]
\caption{\label{be2}Experimental and calculated $B(E2)$ and  $B(M1)$ in W.u. for different
transitions with $e_p$=1.5e, $e_n$=1.1e.}
\begin{center}
\begin{tabular}{rrrrr}
Nucleus    &  Transition                                    &  Expt.            & JUN45 & jj44b   \\       
\hline
$^{87}$Sr  & $B(E2;5/2^{+}_{1} \rightarrow 9/2^{+}_{1})$    & 7.5 (23)          & 10.19 & 12.27   \\   
           & $B(E2;13/2^{+}_{1} \rightarrow 9/2^{+}_{1}$    & 5.5 (17)          & 8.78  &  9.67   \\  
           & $B(E2;5/2^{+}_{2} \rightarrow 9/2^{+}_{1})$    & $0.13_{-13}^{+5}$ & 0.34  &  0.60   \\ 
           & $B(E2;7/2^{+}_{1} \rightarrow 9/2^{+}_{1})$    & 1.9 (5)           & 3.63  &  4.46   \\
           & $B(E2;11/2^{+}_{1} \rightarrow 9/2^{+}_{1})$   & $>$ 2.0           & 4.48  &  7.45   \\ 
\hline
$^{87}$Zr  &$B(E2;7/2^{+}_{1} \rightarrow 9/2^{+}_{1})$     & 3.25 (14)         & 17.68 & 16.48   \\
           &$B(E2;13/2^{+}_{1} \rightarrow 9/2^{+}_{1})$    &  $>$ 0.36         & 22.17 & 38.92   \\
           &$B(E2;21/2^{+}_{1} \rightarrow 17/2^{+}_{1})$   & $>$ 2.19(22)      & 6.26  &  1.82   \\          
\hline
$^{87}$Sr  &$B(M1;7/2^{+}_{1} \rightarrow 9/2^{+}_{1})$     & 0.013 (3)         & 0.021  & 0.0173 \\
           &$B(M1;11/2^{+}_{1} \rightarrow 9/2^{+}_{1})$    & $>$0.013          & 0.0489 & 0.057  \\ 
           &$B(M1;3/2^{+}_{2} \rightarrow 5/2^{+}_{2})$     & 0.09 (4)          & 0.009  & 0.0004 \\
           &$B(M1;3/2^{-}_{1} \rightarrow 1/2^{-}_{1})$     & 0.11 (5)          & 0.246  & 0.181  \\  
\hline
$^{87}$Zr  &$B(M1;7/2^{+}_{1} \rightarrow 9/2^{+}_{1})$     & 0.00095 (4)       & 0.003& 0.0275\\
           &$B(M1;17/2^{-}_{2} \rightarrow 17/2^{-}_{1})$   & ${0.4}_{-4}^{+8}$ & 0.044& 0.084\\
           &$B(M1;19/2^{-}_{2} \rightarrow 17/2^{-}_{1})$   & 0.0074 (21)       & 0.166 & 0.141 \\   
           \hline
\end{tabular}
\end{center}
\end{table}

\begin{table}
\begin{center}
\caption{\label{qm} Electric quadrupole moments, $Q_s$ (in eb), with the two different interactions (the effective
charges $e_p$=1.5e, $e_n$=1.1e are used in the calculation)}
\begin{tabular}{c|c|c|c} 
              &              &$^{87}$Sr    &   $^{87}$Zr   \\ 
\hline
$Q(9/2_1^+)$  & Expt         &  +0.305 (2) &   +0.423 (48) \\ 
              & JUN45        &  +0.349     &   +0.341      \\ 
              & jj44b        &   +0.415    &   +0.649      \\ 
\hline        
$Q(5/2_1^+)$  &   Expt       &  N/A        &  N/A          \\ 
              & JUN45        &  +0.267     & +0.202       \\
              & jj44b        & +0.289      & +0.474        \\ 
\hline         
$Q(7/2_1^+)$  &   Expt       &  N/A        &  N/A          \\ 
              & JUN45        &  +0.0129    & +0.439        \\
              & jj44b        & +0.0237     & +0.484        \\ 
\hline        
$Q(11/2_1^+)$ &   Expt      &  N/A        &  N/A          \\ 
              & JUN45        &  +0.224     & +0.606       \\
              & jj44b        & +0.239      & +0.597        \\ 
\hline        
$Q(13/2_1^+)$ &   Expt       &  N/A        &  N/A          \\ 
              & JUN45        &  +0.510     & +0.583        \\
              & jj44b        & +0.549      & +0.835        \\              
\hline             
\end{tabular}
\end{center}
\end{table}
\begin{table}
\begin{center}
\caption{\label{mm} For magnetic moments  $\mu$ (in $\mu_N$), here $g_s^{\rm eff}$ =0.7$g_s^{\rm free}$ is used in the calculation.}
\begin{tabular}{c|c|c|c} 
                    &            & $^{87}$Sr      & $^{87}$Zr \\ \hline
 $\mu$ ($9/2_1^+$)  &   Expt     &    -1.0928 (7) &  -0.895(9)\\ 
                    &  JUN45     &  -1.0298       &  -0.9401  \\ 
                    &  jj44b     &  -0.9796       &  -0.7816  \\ 
\hline        
 $\mu$($7/2_1^+$)   &  Expt      &   N/A         &     N/A   \\ 
                    & JUN45      &  -0.9988      &   -0.7169 \\ 
                    & jj44b      &  -0.4717      &   -0.5596 \\ 
\hline     
 $\mu$($1/2_1^-$)   & Expt       &  +0.624 (4)   & +0.642 (7)\\
                    & JUN45      &   0.498       &  +0.437   \\
                    & jj44b      &   0.472       &  +0.395   \\
\hline
                    \end{tabular}
\end{center}
\end{table}

\section{Electromagnetic properties}

In Table~\ref{be2}, we have reported experimental versus calculated $B(E2)$ and  $B(M1)$ values in W.u. with different
transitions. All the $B(E2)$ values of $^{87}$Sr are in better agreement with the experiment in both calculations
and $B(E2)$ values of $^{87}$Zr are predicted much larger than in the experiment in both calculations.
We have used the recommended  $e_p$=1.5e, $e_n$=1.1e values of effective charges~\cite{jun45} in the JUN45 interaction.
 Larger neutron effective charge is more reasonable in this mass region which is adopted by  Honma {\it et al.}  \cite{jun45}
from the least-squares fit to the 49 known experimental values of quadrupole moments. With the $(e_p,e_n)$=(1.5,1.1) 
the agreement between theory and experiment is very good. In the present work,
large $B(E2)$ values are due to many nucleons in the valence shells and the agreement may be improved by slightly reducing the effective charges. 
Quality of the magnetic moment agreement with the experimental data is also like quadrupole moments: the magnetic
moments of $^{87}$Sr are better described by the calculations. 

The electric quadrupole and magnetic moments we have listed in Tables~\ref{qm} and~\ref{mm}, respectively.
For the $^{87}$Sr nucleus the $Q(9/2_1^+)$ value is +0.349 eb  according to the calculation with JUN45 interaction, which is
closest to the experimental +0.305(2) eb  value. The jj44b predicts the value larger than experimental one. In the case of $^{87}$Zr, the $Q(9/2_1^+)$ value is lower for JUN45,
while for jj44b it is very large. Though in general quadrupole moments are in excellent agreement with the experimental data,
the agreement still can be improved by reducing effective charges which is reasonable for these nuclei. The shape
of $^{87}$Sr is more spherical than $^{87}$Zr in the $7/2_1^+$ state according to the quadrupole moment values calculated by both  interactions for this state. 

For the calculation of magnetic moments, 
$g_s^{\rm eff}$ =0.7$g_s^{\rm free}$ is used as recommended in the ref.~\cite{jun45}. The results of JUN45
interaction, is in very good agreement with experimental data. In the case of jj44b calculation the predicted value
is slightly lower. Here $g_s^{\rm eff}$ = $g_s^{\rm free}$ may improve the results.

\section{Conclusions}

Motivated by recent experimental results for high-spin states in $^{87}$Sr and $^{87}$Zr, we
performed shell model calculations for these nuclei in $f_{5/2}pg_{9/2}$ model space using JUN45 and
jj44b effective interactions. The results for energy levels and electromagnetic transitions are presented.
The high spin energy levels are described very well by the effective interactions for the full
$f_{5/2}pg_{9/2}$ model space. In general both  effective interactions show very good agreement with the experimental data,
For $^{87}$Zr, the jj44b interaction predicts negative parity states beyond $J \geq 27/2^{-}$ higher in energy.
 The calculated values of quadrupole moment are in good agreement with 
available experimental data.  

\vspace{0.8cm}

{\bf  Acknowledgement}
\vspace{0.2cm}

PCS acknowledges support from faculty initiation grants. VK gratefully acknowledges financial support from CSIR-India.
MJE's work is supported by TWAS-CNPq grant
and grant F2-FA-F177 of Uzbekistan Academy of Sciences.


\newpage
\begin{thebibliography}{99}

\bibitem{Kum}  G. J. Kumbartzki {\it et al.}, Phys. Rev. C {\bf 85}, 044322 (2009).
\bibitem{th1}  J. Skalski, S. Mizutori, and W. Nazarewicz, Nucl. Phys.A {\bf 617}, 282 (1997).
\bibitem{jun45} M. Honma, T. Otsuka, T. Mizusaki and M. Hjorth-Jensen, Phys. Rev. C {\bf 80}, 064323 (2009).
\bibitem{ecl} E. Cl\'ement {\it et al.}, Phys. Rev. Lett. {\bf 116}, 022701 (2016).
\bibitem{th2}  A. Holt, T. Engeland, M. Hjorth-Jensen, and E. Osnes, Phys. Rev. C {\bf 61}, 064318 (2000).
\bibitem{th3}  T. Rzaca-Urban, K. Sieja, W. Urban, F. Nowacki, J. L. Durell, A. G. Smith, and I. Ahmad, Phys. Rev. C {\bf 79}, 024319 (2009).
\bibitem{th4}   K. Sieja, F. Nowacki, K. Langanke, and G. Martinez-Pinedo, Phys. Rev. C {\bf 79}, 064310 (2009).
\bibitem{th5}   R. Rodriguez-Guzman, P. Sarriguren, L. M. Robledo, and S. Perez-Martin, Phys. Lett. B {\bf 691}, 202 (2010).
\bibitem{th6}   Y.-X. Liu, Y. Sun, X.-H. Zhou, Y.-H. Zhang, S.-Y. Yu, Y.-C.Yang, and H. Jin, Nucl. Phys. A {\bf 858}, 11 (2011).
\bibitem{th7}    J. Barea, J. Kotila, and F. Iachello, Phys. Rev. C {\bf 87}, 057301 (2013).
\bibitem{th8}  R. Sahu, V.K.B. Kota and P.C. Srivastava, J. Phys. G: Nucl. Phys. {\bf 40}, 095107 (2013).
\bibitem{jain} A.K.Jain, B.Maheshwari, S.Garg, M.Patial, B.Singh, Nucl. Data Sheets {\bf 128}, 1 (2015).
\bibitem{zr88} S. Saha, R. Palit, J. Sethi, S. Biswas, P. Singh, T. Trivedi, D. Choudhury, P.C. Srivastava, Phys. Rev. C {\bf 89}, 044315 (2014).
\bibitem{zr89} S. Saha, R. Palit, J. Sethi, T. Trivedi, P.C. Srivastava {\it et al.}, Phys.Rev. C {\bf 86}, 034315 (2012).
\bibitem{sr85} S. Kumar, N. Kumar, S. Mandal, S.C. Pancholi, P.C. Srivastava, A.K. Jain {\it et al.}, Phys. Rev. C {\bf 90}, 024315 (2014).
\bibitem{sr86} N. Kumar, S. Kumar, V. Kumar, S.K. Mandal, R. Palit, S. Saha, J. Sethi, T. Trivedi, S.C. Pancholi, and P.C. Srivastava, Nucl. Phys. A {\bf 955}, 1 (2016).
\bibitem{sr871} S E Arnell, A Nilsson, O Stankiewicz, Nucl. Phys. A, {\bf 241}, 109 (1975). 
\bibitem{sr872} L P Ekstr\"om  {\it et al.}, J. Phys. G: Nucl. Phys. {\bf 7}, 85 (1981).
\bibitem{sr873} Li Hong-Wei  {\it et al.},  Chin. Phys. C {\bf 38}, 074004 (2014).
\bibitem{zr871} Guang-yi Zhao {\it et al.},  Chin. Phys. Lett. {\bf 16}, 345 (1999).
\bibitem{jj44b} B.A. Brown and A.F. Lisetskiy (unpublished); see also endnote
(28) in B. Cheal {\it et al.}, Phys. Rev. Lett. {\bf 104}, 252502 (2010).
\bibitem{antoine} E. Caurier,  G. Mart\'inez-Pinedo , F. Nowacki, A. Poves,
and A. P. Zuker, Rev.\ Mod.\ Phys. {\bf77} (2005) 427.
\bibitem{nndc}www.nndc.bnl.gov/ensdf
\end{thebibliography}
\end{document}